\documentclass[onecolumn,amsmath,amssymb,superscriptaddress,nofootinbib,11pt,a4paper]{revtex4} 

\usepackage[colorlinks,linkcolor=red]{hyperref}

\pdfoutput=1
\usepackage[T1]{fontenc}
\usepackage{CJK}
\usepackage{xcolor}
\usepackage{amsfonts}
\usepackage{epstopdf}
\usepackage{wrapfig} 
\usepackage{subfigure}
\usepackage{graphicx}  
\usepackage{dcolumn}   
\usepackage{bm}
\usepackage{float}
\newcommand{\be}{\begin{equation}}
\newcommand{\ee}{\end{equation}}
 \newcommand{\bea}{\begin{eqnarray}}
 \newcommand{\ena}{\end{eqnarray}}

\begin{document}
\title{Thermodynamics and weak cosmic censorship conjecture in the  Kerr-AdS black hole }

\author{Xiao-Xiong Zeng,}\email{xxzengphysics@163.com}
\affiliation{School of Material Science and Engineering, Chongqing Jiaotong University, Chongqing 400074, China}
\author{Hai-Qing Zhang}\email{hqzhang@buaa.edu.cn}
\affiliation{Center for Gravitational Physics, Department of Space Science and International Research Institute of Multidisciplinary Science, Beihang University, Beijing 100191, China}

\begin{abstract}
{We investigate the laws of thermodynamics and weak cosmic censorship conjecture in the normal and extended phase space of a Kerr-AdS black hole by analyzing the energy-momentum relation of the absorbed fermion dropping into the horizon. In the normal phase space, the first law, second law as well as the weak cosmic censorship conjecture are found to be valid in all the initial states of the black hole. However, in the extended phase space, although the first law and weak cosmic censorship conjecture are still valid, the second law becomes more subtle. We find that the validity or violation of the second law depends on the spin parameter, the radius of the AdS spacetime, and their variations. In addition, in the extended phase space, the configurations of the extremal and near-extremal Kerr-AdS black holes   are unchanged  as the fermions are absorbed since the  final and  initial states are the same.}
\end{abstract}

\maketitle
 \newpage
\section{Introduction}


Due to the pioneering work of Hawking \cite{Hawking:1974sw},  we  know that a black hole is not only a celestial body, but also a thermodynamic system. 
Just like the ordinary thermodynamic systems,  there are also four laws for the black hole thermodynamic systems \cite{Hawking:1976de}, which are valid near the horizon.
The importance of thermodynamics of black holes is that it
provides fundamental relations between theories such as gravity,
thermodynamics and quantum field theory.

Thermodynamics in the AdS space is more prevalent for the existence of a cosmological parameter. Initially, the cosmological parameter in the thermodynamic system is treated as a constant. However, more and more evidences have shown that this parameter may be considered as a dynamical variable \cite{Henneaux:1985tv}. Moreover, Ref. \cite{Caldarelli:1999xj} put forward to regarding it as a state function. In particular, Refs. \cite{Dolan:2010ha,Cvetic:2010jb,Kastor:2009wy} proposed that the cosmological constant   can be regarded as the pressure   of the thermodynamic system, while its thermodynamic conjugate variable can be defined naturally as volume. The thermodynamic phase space thus is extended.  In the extended phase space, the  Smarr relation derived by the
scaling method  is also valid  besides the first law of thermodynamics. The mass, of course, is not the internal energy  but rather the enthalpy of the thermodynamic system. The emergent property in the extended phase space is that the phase structures  of a black holes are more abundant, and one can observe the van der Waals-like phase transition \cite{Bhattacharya:2017nru,Zou:2017juz,Zeng:2015wtt,Xu:2015rfa,Cai:2013qga,Hu:2018qsy}, reentrant phase transitions \cite{Dehyadegari:2017hvd,Zhang:2017lhl,Kubiznak:2016qmn,Altamirano:2013ane}.

Another reason for the importance of the thermodynamics in  AdS space arises from the AdS/CFT correspondence \cite{Maldacena:1997re,Gubser:1998bc,Witten:1998qj,Aharony:1999ti}, which
states that the gravity theory in $d$-dimensional AdS spacetime is dual to the conformal field theory in  $(d-1)$-dimension. 
Various holographic applications have been proposed in the last two decades, for example, the holographic superconductors \cite{Hartnoll:2008vx,Cai:2015cya,Cai:2012sk,Cai:2012nm}, holographic thermalizations \cite{Balasubramanian1,Gb,Gb2,Zeng:2014xpa},  holographic Fermi/non-Fermi   liquids \cite{Lee,Liu,Cubrovic}, and so on.

For the general relativity, if singularities are naked, causality in the spacetimes may break down, and physics will lose its predictive power.  Penrose  conjectured that  the
singularity will not be naked for it is always covered  by an event horizon \cite{Penrose}, which is the so-called `weak cosmic censorship conjecture'.  Since there are no general  ways to
 prove  this  conjecture, we thus should investigate each type of black holes.
Wald proposed firstly a gedanken experiment to check this conjecture by throwing a particle  into the Kerr-Newman black hole, and he found that the weak cosmic censorship conjecture was valid for the extremal Kerr-Newman  black holes
\cite{Wald:1974ge}. However, for the  near-extremal Reissner-Nordstr\"{o}m \cite{Hubeny:1998ga} black hole  and near-extremal Kerr black hole\cite{Jacobson:2009kt}, this conjecture were found to be violated.
Later it was pointed out  that the invalidity of  the weak cosmic censorship conjecture in \cite{Hubeny:1998ga,Jacobson:2009kt}   was that the   backreaction and self-force effects are neglected\cite{Barausse:2010ka}.
 So far, the weak cosmic censorship conjecture has been checked in various black holes with and without the  self-force effect by
adding a particle \cite{Colleoni:2015afa,BouhmadiLopez:2010vc,Hod:2016hqx,Natario:2016bay,
  Horowitz:2016ezu,Duztas:2016xfg,Gwak:2017icn,
Gao:2012ca,Rocha:2014jma,Rocha:2011wp,Chen:2018yah,Wald2018,Sorce:2017dst,Ge:2017vun,An:2017phb,Crisford2017,Gwak:2016gwj}.

As mentioned above, there are many works to investigate the thermodynamics of black holes in the extended phase space, however all of them focus only on the first law of thermodynamics. There is little work to discuss the second law of thermodynamics and the weak cosmic censorship conjecture  in the extended phase space.
Recently, Ref. \cite{Gwak:2017kkt}
suggested that the first law, second law as well as the weak cosmic censorship conjecture  can be  investigated  in the  extended phase space under charged particles absorption.
Their results showed that though the first law and the weak cosmic censorship conjecture
were valid,
the second law   was violated  for the near-extremal Reissner-Nordstr\"{o}m-AdS black holes.
 Subsequently,  the idea in   \cite{Gwak:2017kkt} was extended to Born-Infeld-anti-de Sitter black hole \cite{Zeng:2019jta}, torus-like black hole \cite{Han:2019kjr} and charged fermions absorption \cite{Zeng:2019jrh,Chen:2019pdj,Han:2019lfs}. In this paper, we are going to investigate the laws of thermodynamics and weak cosmic censorship conjecture in the Kerr-AdS black hole. Different form previous investigations, the topology of this black hole is axisymmetric.
In fact, Refs.\cite{Gwak:2018tmy,Gwak:2015fsa} have investigated the laws of thermodynamics  of this black hole in the normal phase space. In order to get the first law as well as the second law,  a reference energy of the particle in the asymptotic region was imposed.  In this paper, we will explore whether the first law and second law  may be obtained by the energy-momentum relation of the particles without any imposition.  In addition, the laws of thermodynamics and  weak cosmic censorship conjecture of the rotating black holes in the extended phase space are also explored. 

This paper is arranged as follows. In section~\ref{sec2}, we briefly review  the thermodynamics of the Kerr-AdS black hole, and then study the dynamics  of the spinning fermions in this black hole. In section ~\ref{sec3}, we investigate the first law, the  second law  as well as the weak cosmic censorship conjecture in the normal phase space. The law of thermodynamics and  weak cosmic censorship conjecture in the extended phase space are studied in section ~\ref{sec4}.
Section~\ref{sec5} is devoted to our conclusions.
Throughout this paper, we set the gravitational constant $G$ and the light velocity $c$ to be one.

\section{Energy and momentum of the spinning fermions}
\label{sec2}

\subsection{Brief review of the  Kerr-AdS black holes}

Line element of the Kerr-AdS black hole  in Boyer-Lindquist coordinates can be written as \cite{Carter:1968ks}
\bea \label{metric}
 ds^{2}=-\frac{\Delta}{\rho^{2}}\bigg(dt-\frac{a\sin^{2}\theta}{\Xi}d\varphi\bigg)^{2}
        +\frac{\rho^{2}}{\Delta}dr^{2}+\frac{\rho^{2}}{\Delta_{\theta}}d\theta^{2}
        +\frac{\Delta_{\theta}\sin^{2}\theta}{\rho^{2}}\bigg(adt-\frac{r^{2}+a^{2}}{\Xi}d\varphi\bigg)^{2},
\ena
with the metric functions given by
\bea
 \Delta&=& (r^{2}+a^{2})(1+r^{2}/l^{2})-2 mr,\label{delta}~~~
 \rho^{2}= r^{2}+a^{2}\cos^{2}\theta,\\
 \Xi&=&1-\frac{a^{2}}{l^{2}},~~~
\Delta_{\theta}= 1-a^{2}/l^{2}\cos^{2}\theta,
\ena
in which,  $m$, $a$ and $l$ are the black hole mass parameter, spin parameter and the AdS radius respectively. Cosmological constant $\Lambda$ is related to the AdS radius as $\Lambda=-3/l^2$.

Matter  near the event horizon will be dragged inevitably by the gravitational field with an azimuthal angular velocity defined by
\be \label{av}
\bar{\Omega} =\frac{a\Xi\big(\Delta_{\theta}(r ^{2}+a^{2})-\Delta\big)}{\Delta_{\theta}(r ^{2}+a^{2})^2-\Delta a^2 \sin^2\theta}.
\ee
The angular velocities at the event  horizon and infinity can be readily obtained from Eq.(\ref{av}) as
\be
\bar{\Omega}_h=\frac{a\Xi}{r_{  h}^{2}+a^{2}},~~~
\bar{\Omega}_{\infty}=-\frac{a}{l^2},
\ee
where $r_h$ is position of the event horizon. It should be noted that the angular velocity at infinity does not vanish, which is distinct from that in the asymptotic flat case. Thus, the net angular velocity  that contributes to the thermodynamics is given by \cite{Caldarelli:1999xj}
 \be
\Omega_h=\bar{\Omega}_h-\bar{\Omega}_{\infty} =\frac{a\Xi}{r_{  h}^{2}+a^{2}}+\frac{a}{l^{2}}.\\ \label{omega}
\ee
Hawking temperature and Bekenstein-Hawking entropy of the Kerr-AdS black holes  are
\bea \label{T}
 T_h=\frac{r_h}{4\pi(r_h^{2}+a^{2})}
   \left(1+\frac{a^{2}}{l^{2}}+3\frac{r_h^{2}}{l^{2}}-\frac{a^{2}}{r_h^{2}}\right),~~~
 S=\frac{\pi (r_h^{2}+a^{2})}{\Xi}. \label{S}
\ena
 The black hole mass $M$ and angular momentum $J$ can be expressed by the mass parameter $m$ and spin parameter $a$ as $M=\frac{m}{\Xi^{2}}, J=\frac{am}{\Xi^{2}}$, respecitvely.
Therefore, we get the first law of thermodynamics
\be \label{firstlaw}
dM=T_h dS+ \Omega_h dJ.
\ee
    In Eq.(\ref{firstlaw}), the cosmological parameter is treated as a constant, and the resulting phase space of the thermodynamics is called `normal phase space'. Recent studies have shown that the normal phase space can be extended by treating the cosmological parameter  as an extensive  variable, leading to an `extended phase space' \cite{Dolan:2010ha,Cvetic:2010jb,Kastor:2009wy}. Thus, the first law of thermodynamics in the extended phase space can be written as \cite{Altamirano:2013uqa,Cheng:2016bpx}
\be \label{firstlaw1}
dM=T_h dS+ \Omega_h dJ+VdP,
\ee
in which $P$ and $V$ are respectively the pressure and volume of the system, defined by
\bea \label{pp}
P=\frac{3}{8 \pi  l^2},~~
V=\frac{2 \pi  \left(a^2+r_h^2\right) \left(a^2 l^2-a^2 r_h^2+2 l^2 r_h^2\right)}{3 l^2 \Xi ^2 r_h}.
\ena
Based on the first law in the extended phase space, the phase structures of the  Kerr-AdS black holes have been investigated extensively. It was shown that in the $P-V$ plane,  there were phase transitions between the small black holes and large black holes \cite{Altamirano:2013uqa,Cheng:2016bpx}, which is similar to the  van der
Waals  phase transition.

\subsection{Motion of the  spinning fermions in the  Kerr-AdS
black holes}

When particles are absorbed by spherical symmetric black holes, it has been confirmed that they satisfy the first law of thermodynamics as the energy-momentum relations of particles  are employed. We will adopt the similar strategy to investigate the laws of thermodynamics of Kerr-AdS black holes, when the absorbed particles are spinning fermions. The Dirac equation now is
$i\gamma ^{\mu }\left( \partial _{\mu }+\Omega _{\mu } \right) \psi -\frac{ u }{\hbar }\psi =0$, 
in which $u$ is the rest mass of the fermion,  while  $\Omega _\mu = \frac{i}{2}\Gamma _\mu ^{\alpha \beta
} \prod _{\alpha \beta } $, $\prod _{\alpha \beta } =
\frac{i}{4}\left[ {\gamma ^\alpha ,\gamma ^\beta } \right]$,  and Dirac matrices $\gamma
^\mu $ satisfy $\left\{ {\gamma ^\mu ,\gamma ^\nu }
\right\} = 2g^{\mu \nu }$.

In order to  investigate the motion of the fermions in the rotating black holes, one should carry out the dragging coordinate transformation with the definition  $d\phi=d\varphi-\bar{\Omega} dt$.
In this case, the metric in  Eq.(\ref{metric})  becomes
\be\label{newmetric}
 ds^{2}=-F(r)dt^2+\frac{1}{G(r)}dr^2+H(r)d\theta^2+K(r)d\phi^2,
\ee
in which
\bea
F(r)&=&\frac{ \Delta \Delta_{\theta} \rho^2}{\Delta_{\theta}(r ^{2}+a^{2})^2-\Delta a^2 \sin^2\theta}, ~~~G(r)=\frac{\Delta}{\rho^2},\\
H(r)&=&\frac{\rho^2}{\Delta_{\theta}},~~~ K(r)=\frac{\sin^2\theta}{\rho^2 \Xi^2}\big((r ^{2}+a^{2})^2\Delta_{\theta}-\Delta a^2 \sin^2\theta\big).
\ena
From Eq.(\ref{newmetric}), we know that the event horizon  and
infinite red-shift surface coincide. This coincidence is important to investigate the tunneling radiation of the particles, for in this case Landau's condition of the
coordinate clock synchronization is satisfied.

To solve the Dirac equation in Eq.(\ref{newmetric}), we
choose $\gamma ^\mu $ matrices as \cite{Chen:2008ge}
\bea \label{gamma}
\gamma ^{t}=\frac{1}{\sqrt{F}}\left(
\begin{array}{cc}
i & 0 \\
0 & -i
\end{array}%
\right) ,~\gamma ^{r}= \sqrt{G} \left(
\begin{array}{cc}
0 & \sigma ^{3}\\
\sigma ^{3} & 0%
\end{array}%
\right),
~\gamma ^{\theta}=\frac{1}{\sqrt{H}}\left(
\begin{array}{cc}
0 & \sigma ^{1}\\
\sigma ^{1} & -1%
\end{array}
\right),~\gamma ^{\phi}=\frac{1}{\sqrt{K}}\left(
\begin{array}{cc}
0 & \sigma ^{2} \\
\sigma ^{2} & 0%
\end{array}
\right).
\ena%
in which $\sigma ^i $ are the Pauli matrices. 
For a fermion with spin $1/2$,  the wave function have spin up and down state. In this paper, we are interested in the
spin up state since the spin down  case is similar.
The wave function  for the spin up state is supposed to be
\begin{equation}~~~~~~~~~~~~~~~
\label{wave}
\psi  = \left(
{{\begin{array}{*{20}c}
 {C\left( {t,r,\theta, \phi } \right)} \hfill \\
 0\\
 {D\left( {t,r,\theta,\phi } \right)} \hfill \\
 0
\end{array} }} \right)\exp \left( {\frac{i}{\hbar }I \left(
{t,r,\theta, \phi } \right)} \right).
\end{equation}
Inserting Eq.(\ref{wave}) and  Eq.(\ref{gamma}) into  the Dirac equation,  and applying
the WKB approximation,
we are led to%

\begin{eqnarray}\label{radial1}
-\bigg( \frac{i C}{\sqrt{F%
}}\partial _{t}I+D\sqrt{G}\partial _{r}I\bigg)+u C=0,~~~
\bigg( \frac{i D}{\sqrt{F%
}}\partial _{t}I-C\sqrt{G}\partial _{r}I\bigg)+u D=0, \\
\label{radial2}
-D\bigg( \frac{1}{\sqrt{H%
}}\partial _{\theta}I+ \frac{i}{\sqrt{K%
}}\partial _{\phi}I\bigg)=0,~~~
-C\bigg( \frac{1}{\sqrt{H%
}}\partial _{\theta}I+ \frac{i}{\sqrt{K%
}}\partial _{\phi}I\bigg)=0,
\end{eqnarray}
in which we only take the leading order of $\hbar$. In order to solve these equations, one should separate the quantity $I$.
According to the symmetries of the spacetime, the quantity $I$
in the dragging coordinate system can be rewritten as \cite{Chen:2008ge}
\begin{equation} \label{eq:I}
I=-(\omega-j \Omega_h )t +I\left( r\right)+L\phi+\Theta (\theta),
\end{equation}%
where $\omega $ and $j$ are respectively fermion's energy and angular momentum
 measured by the observer at
the infinity. Substituting Eq.(\ref{eq:I}) into Eqs.(\ref{radial1}) and  (\ref{radial2}), we obtain
\begin{equation}\label{radial3}
C\bigg(u+ \frac{i}{\sqrt{F%
}} (\omega -j \Omega_h)\bigg) -D\sqrt{G}\partial _{r}I=0,~~~
-C\sqrt{G}\partial _{r}I+D\bigg(u- \frac{i}{\sqrt{F%
}} (\omega -j \Omega_h)\bigg)=0. 
\ee
 Eqs.(\ref{radial3}) 
 have non-trivial solutions for $C$ and $D$ only
if the determinant of the coefficient matrix vanishes. From this condition, we can get $\partial _{r}I$, and further the radial momentum $p^r=g^{rr}\partial _{r}I$,
\be \label{prr}
p^r =\pm \sqrt{\frac{G}{F}}\sqrt{\left( \omega -j \Omega_h\right) ^{2}+\mu ^{2}F  }.
\ee
The laws of thermodynamics should be studied near horizon region. Thus,  Eq.(\ref{prr}) is simplified as
\begin{equation}\label{emr}
\omega=\frac{\rho_h^2 }{a^2+r_h^2} |p^r_h|+j \Omega_h,
\end{equation}
in which the lower index $h$ represents near horizon region, with $\rho_h^2=a^2 \cos^2\theta+r_h^2$.
  From Eq.(\ref{emr}), we know that when $\omega< j \Omega_h$, the
supperradiation occurs and the energy of the black hole flows out of the horizon, since the sign in front of  $|p^r_h|$  is  negative. In this paper, we assume that the supperradiation does not
occur, which implies $\omega > j \Omega_h$ and  the sign in front of  $|p^r_h|$ is  positive.

\section{Thermodynamics and weak  cosmic  censorship  conjecture in the normal phase space }\label{sec3}
We will employ the energy-momentum relation derived in Eq.(\ref{emr}) to study the first, the second laws of thermodynamics and the  weak  cosmic  censorship  conjecture in the normal phase space of Kerr-AdS black hole in this section. 

\subsection{Laws of thermodynamics in the normal phase space}
We assume that the absorbed fermion has energy $\omega$ and angular momentum $j$. As it drops into  the black hole, the internal energy $M$ and angular momentum $J$ of the black hole would change according to
\begin{equation}\label{dmdj}
\omega=dM,\quad j=dJ,
\end{equation}
Due to the first law of thermodynamics, Eq.(\ref{emr}) can be rewritten as
\begin{equation}\label{newemr}
dM=j \Omega_h+\frac{\rho_h^2 }{a^2+r_h^2}p^r_h.
\end{equation}
In the normal phase space, the initial state of
the black hole is represented by $(m, a, r_h)$, while the final state is represented by $(m+dm, a+da, r_h+dr_h)$. That is, the absorbed fermions will change the configurations of the black holes. However, the horizon is always determined  by the function $\Delta(r)$, i.e., $\Delta(r_h+dr_h)=0$ always holds. Thus,  the change of the horizon    should satisfy
\begin{equation}\label{ddelta}
d\Delta_{h}=\frac{\partial \Delta_{h}}{\partial m}dm+\frac{\partial \Delta_{h}}{\partial a}da+\frac{\partial\Delta_{h}}{\partial r_{h}}dr_{h}=0.~~~~
\end{equation}
In addition, by considering $M=\frac{m}{\Xi^{2}}$,
 Eq.(\ref{newemr}) can be rewritten as
\be \label{ep}
 \frac{\partial }{\partial a}\frac{m}{\Xi ^2}da+  \frac{\partial }{\partial m}\frac{m}{\Xi ^2}dm-\left(\frac{a \Xi }{a^2+r_h^2}+\frac{a}{l^2}\right) \left( \frac{\partial }{\partial a}\frac{a m}{\Xi ^2}da+ \frac{\partial }{\partial m}\frac{a m}{\Xi ^2}dm\right) -\frac{\rho_h^2 }{a^2+r_h^2}p^r_h =0.
\ee
After solving  Eq.(\ref{ep}), we reach
 \be\label{dm}
  dm=\frac{a  l^2 m \left(l^2-3 r_h^2\right)  da  }{l^2 r_h^2 \left(l^2-a^2\right)}+\frac{\rho_h^2\big[a^6 p^r_h +a^4 p^r_h  \left(r_h^2-2 l^2\right)+a^2 l^2 p^r_h  \left(l^2-2 r_h^2\right)+l^4 p^r_h   r_h^2\big]}{\big(a^2+r_h^2\big)\big[l^2 r_h^2 \left(l^2-a^2\right)\big]}.
 \ee
Inserting Eq.(\ref{dm}) into Eq.(\ref{ddelta}), we arrive at
 \be\label{eq:variables0233}
dr_{h}=\frac{a  l^2  da  }{a^2 r_h-l^2 r_h}+\frac{a  r_h  da  }{a^2-l^2}+\frac{2 p^r_h  \big(a-l\big) \big(a+l\big) \left(a^2+r_h^2\right)\rho_h^2 }{\big(a^2+r_h^2\big)\big[a^2 l^2-r^2 \left(a^2+l^2\right)-3 r_h^4\big]}.
\ee
Based on the above formula,  one can  readily obtain  the variation of entropy by making use of Eq.(\ref{S}), which is
\begin{equation}\label{eq:s0233}
dS=\frac{\partial S}{\partial a}   da   + \frac{\partial S}{\partial r_h} {dr_h}=\frac{4 \pi  l^2  r_h \left(a^2+r_h^2\right)\rho_h^2 p^r_h}{(a^2+r_h^2)[r_h^2 \left(a^2+l^2\right)-a^2 l^2+3 r_h^4]}.
\end{equation}
With the help of Eqs.(\ref{T}) and (\ref{eq:s0233}), we can get the following  relation
 \be \label{prts}
T_h dS=\frac{\rho_h^2 }{a^2+r_h^2}p^r_h.
 \ee
Therefore, the internal energy in Eq.(\ref{newemr}) can be rewritten as
\begin{equation} \label{eq:dm1}
dM=T_h dS+ \Omega_h dJ,
\end{equation}
which is  consistent with Eq.(\ref{firstlaw}). Therefore, we see that as a spinning  fermion drops into the Kerr-AdS black hole, the first law of thermodynamics
is valid in the normal phase space.


For the extremal  black holes, the inner horizon and outer horizon coincide and the temperature vanishes at the horizon.
  From Eq.(\ref{T}), we can get the radius of the extremal black hole as
 \be \label{er}
 r_{extreme}=\frac{\sqrt{-a^2+\sqrt{a^4+14 a^2 l^2+l^4}-l^2}}{\sqrt{6}}.
 \ee
The second law of thermodynamics for the extremal black holes is meaningless for the temperature vanishes. In fact, from Eq.(\ref{prts}), we know that the variation of the entropy is divergent in this case.
For the non-extremal  black hole, the temperature is larger than zero for $r>r_{extreme}$, thus $dS$ in  Eq.(\ref{eq:s0233}) is positive, implying that the second law of thermodynamics is also valid in the normal phase space for the non-extremal Kerr-AdS black holes.

\subsection{Weak  cosmic  censorship  conjecture in the normal phase space }

The validity of the weak cosmic censorship can be tested by computing the minimum value of the
function  $\Delta(r)$ after the absorption of fermions. For the Kerr-AdS black holes, we label the radial position of the minimum of $\Delta(r)$ as $r_{l}$.    At $r_{l}$,
 the following relations always satisfy
\bea \label{condition1}
\Delta(r)|_{r=r_{l}}\equiv \Delta_{l}=\lambda\leq 0,~~~
\partial_{r}\Delta(r)|_{r=r_{l}}\equiv \Delta'_{l}=0.
\ena
 For the extremal black holes, $\lambda=0$, the horizon and the location of the minimum value are coincident. For the near-extremal black holes, $\lambda$ is a small quantity, the location of the minimum value is located between the inner horizon and outer horizon.  As a fermion with mass $\omega$ and angular momentum $j$ drops into the black hole, the mass and angular momentum of the black hole  will  increase as $M+\omega, J+j$ respectively. Correspondingly, the location of
the minimum value and event horizon change into $ r_{l}+dr_{l}$ and $r_{h}+dr_{h}$, respectively. 
Thus, at $r_l+dr_l$, we have
\bea \label{move1}
\Delta(r_{l}+dr_{l})=\Delta_{l}+d\Delta_{l}=\lambda+\left(\frac{\partial \Delta_{l}}{\partial m}dm+\frac{\partial \Delta_{l}}{\partial a}da\right),
\ena
where we have used   Eq.(\ref{condition1}). For the case of $\Delta(r_{l}+dr_{l})>0$, there is no horizon while for the case of $\Delta(r_{l}+dr_{l})\leq0$, there are always horizons. In the following we will focus on  finding the final form of   Eq.(\ref{move1}).
For simplicity, we consider the extremal black holes, since the horizon is located at $r_l$ and $\Delta_{l}=0$. In this case, Eq.(\ref{dm}) can be adopted, thus we reach
\bea
 d\Delta_{l}&=&\frac{-2 a    da   \big(l^2 r_l \left(a^2-r_l (3 m+r_l)\right)+a^2 r_l^3+l^4 (m-r_l)\big)}{l^2 r_l \left(l^2-a^2\right)} 
 - \frac{2  \left(a^2-l^2\right)^2 \left(a^2+r_l^2\right)(a^2\cos^2\theta+r_l^2)}{l^2 r_l \left(l^2-a^2\right)(a^2+r_l^2)}. ~~~~~\label{move2}
\ena
In addition, from Eq.(\ref{delta}) the mass of the black hole can be written as
\be \label{ml}
m= \frac{\left(a^2+r_l^2\right) \left(l^2+r_l^2\right)}{2 l^2 r_l}.
\ee
Substituting Eqs.(\ref{ml}) and  (\ref{er}) into Eq.(\ref{move2}), we finally get
\be
d\Delta_{l}=\frac{\sqrt{\frac{2}{3}} (a-l) (a+l) \left(a^2 \left(6 \cos^2\theta-1\right)+\sqrt{a^4+14 a^2 l^2+l^4}-l^2\right)}{l^2 \sqrt{-a^2+\sqrt{a^4+14 a^2 l^2+l^4}-l^2}}.
\ee
Values of  $d\Delta_l$ are shown in Figure (\ref{fig1}) for various $l$'s. It is obvious  that for a given value of $l$,   the values of  $d\Delta_l$ can be positive or negative. Take the example of $l=0.7$, we see that for the case of $a<l$, $d\Delta_l$ is negative; while for the case of $a>l$, $d\Delta_l$ is positive. For other values of $l$'s, we can observe similar behaviors of $d\Delta_l$. Therefore, we conclude that for the case of $a<l$,  the weak cosmic censorship conjecture is valid; while for the case of $a>l$,  the weak
cosmic censorship conjecture is violated. Our results are consistent with those in Ref. \cite{Caldarelli:1999xj}, in which the authors found that the Kerr-AdS solution was valid only for $a<l$ since the singularity will be naked for $a\geq l$. \footnote{Please note that in higher dimensional Kerr-AdS solutions,  $a>L$ is also possible. Please refer to \cite{McInnes:2019rtj}. }
\begin{figure}[h]
\centering
\includegraphics[scale=0.45]{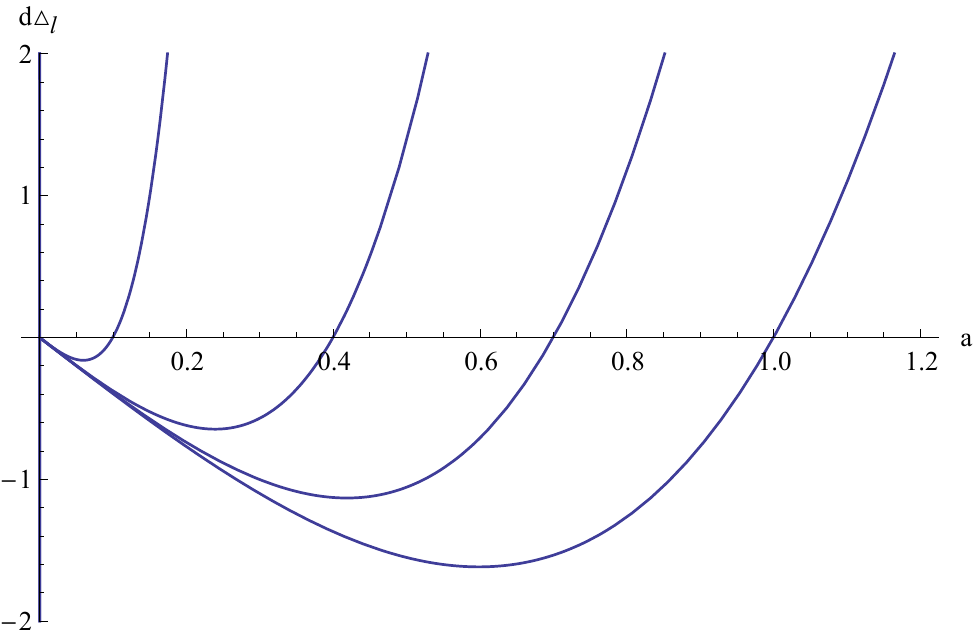}
 \caption{\small The values  of $d\Delta_l$ for the case of $p^r_h = 1, \theta = 0$. Curves from left to  right correspond to $l$ varying from  0.1 to 1 with
 step 0.3.   } \label{fig1}
\end{figure}
\section{Thermodynamics and weak  cosmic  censorship  conjecture in the  extended  phase space }
\label{sec4}

\subsection{Laws of thermodynamics in the extended  phase space}

In the extended phase space, the mass is no longer related to the internal energy but rather the  enthalpy of the thermodynamic system.  The relation between the enthalpy $M$, internal energy $U$, pressure $P$ and volume $V$ are
\begin{align}
M=U+PV.  \label{enthalpy}
\end{align}
As the fermions  are absorbed by the black holes, the energy and angular momentum are supposed to be conserved. Thus, the energy and angular momentum of the spinning fermion equal to the varied energy and angular momentum of the black hole, which leads to
\begin{equation} \label{0pvep}
\omega=dU=d(M-PV),\quad j=dJ,
\end{equation}
where we have used Eq.(\ref{dmdj}). In this case, Eq.(\ref{emr}) changes correspondingly as
\begin{equation}\label{pvnewemr}
d(M-PV)=j \Omega_h+\frac{\rho_h^2 }{a^2+r_h^2}p^r_h.
\end{equation}
    Considering $M=\frac{m}{\Xi^{2}}$,
 Eq.(\ref{pvnewemr}) can be rewritten as
\bea \label{pvep}
  &&\frac{\partial }{\partial a}\frac{m}{\Xi ^2}da+ \frac{\partial }{\partial l}\frac{m}{\Xi ^2}dl+ \frac{\partial }{\partial m}\frac{m}{\Xi ^2}dm -  V \frac{\partial P}{\partial l}dl-P Y-p^r_h  X\nonumber\\
&&-\left(\frac{a \Xi }{a^2+r_h^2}+\frac{a}{l^2}\right) \left(\frac{\partial }{\partial a}\frac{a m}{\Xi ^2}da+  \frac{\partial }{\partial l}\frac{a m}{\Xi ^2}dl+ \frac{\partial }{\partial m}\frac{a m}{\Xi ^2}dm\right)=0,
\ena
in which
\bea
&&Y\equiv dV= \frac{\partial V}{\partial a}   da  + \frac{\partial V}{\partial l}   dl  + \frac{\partial V}{\partial r_h} {dr_h},  \label{dv}~~~ X=\frac{a^2\cos^2\theta+r_h^2}{a^2+r_h^2}.
\ena
By solving  Eq.(\ref{pvep}), we   obtain
  \bea \label{pvdm11}
 dm=\frac{U_1+U_2+U_3+U_4} {8 \pi  l^4 r_h^3 \left(l^2-a^2\right)},
 \ena
where
\bea
U_1&=&8 \pi  a l^4 m r_h \left(l^2-3 r_h^2\right)   da  +l^3 r_h^3 \left(-8 \pi  r_h^3 dl   +8 \pi  l^3 p^r_h  X+3 l Y\right), \nonumber\\
U_2&=& a^6 \left(4 \pi  l \left(r_h^2-l^2\right)   dl   +r_h \left(8 \pi  l^2 p^r_h  X+3 Y\right)\right),\nonumber\\
U_3&=&a^2 l r_h \left(4 \pi r_h^2 \left(l^2 (8 m-5 r)+r_h^3\right)   dl  +l \left(l^2-2 r_h^2\right)\right),\nonumber\\
U_4&=&
a^4 r_h \left(r_h^2-2 l^2\right) \left(8 \pi  l r_h   dl   +8 \pi  l^2 p^r_h  X+3 Y\right).\nonumber
 \ena
In the extended phase space, the AdS radius is also a parameter of the black hole. Thus, the initial state should be characterized  by $(m, a, r_h, l)$. As a fermion drops into the black hole, the final configuration of the black hole becomes $(m+dm, a+da, r_h+dr_h,l+dl)$. At the initial state and final state, the horizons are determined by
$\Delta(r_h)=0$ and $\Delta(r_h+dr_h)=0$ respectively, which implies
\begin{equation}\label{pvddelta1}
d\Delta_{h}=\frac{\partial \Delta_{h}}{\partial m}dm+\frac{\partial \Delta_{h}}{\partial a}da+\frac{\partial\Delta_{h}}{\partial r_{h}}dr_{h}+\frac{\partial\Delta_{h}}{\partial l}dl=0.~~~~
\end{equation}
Substituting Eq.(\ref{pvdm11}) into Eq.(\ref{pvddelta1}), we can solve  $dr_{h}$  directly as
 \bea\label{pvdr}
dr_{h}=\frac{V_1+V_2+V_3+V_4}{4 \pi  l^2 r_h \left(l^2-a^2\right) \left(a^2 \left(l^2-r_h^2\right)-r_h^2 \left(l^2+3 r_h^2\right)\right)},
\ena
where
\bea
V_1&=& 4 \pi  a l^2 r_h^2 \left(l^4+4 l^2 r_h^2+3 r_h^4\right)   da   -l^4 r_h^3 \left(8 \pi  l^2 p^r_h  X+3 Y\right), \nonumber \\
V_2&=&
a^6 \left(4 \pi  l \left(l^2-r_h^2\right)  dl  -r_h \left(8 \pi  l^2 P^r_h  X+3 Y\right)\right),\nonumber \\
V_3&=&4 \pi  a^3 l^2 \left(r_h^4-l^4\right)  da  -a^2 l r_h \left(4 \pi  r_h^3 \left(l^2+3 r_h^2\right)  dl  +l \left(l^2-2 r_h^2\right) \left(8 \pi  l^2 p^r_h  X+3 Y\right)\right),\nonumber \\
V_4&=&a^4 \left(\left(2 l^2 r_h-r_h^3\right) \left(8 \pi  l^2 p^r_h  X+3 Y\right)-16 \pi  l r_h^4  dl   \right).
\ena
From Eq.(\ref{S}),    the variation of entropy can be written as
\begin{equation}\label{pvds}
dS= \frac{\partial S}{\partial a}   da  + \frac{\partial S}{\partial l}   dl  + \frac{\partial S}{\partial r_h} {dr_h}.
\end{equation}
With the help of Eqs.(\ref{T}), (\ref{pp}),  (\ref{dv}), and (\ref{pvds}), we find that,
\be \label{pvpr}
T_h dS-PdV=\frac{\rho_h^2 }{a^2+r_h^2}p^r_h.
\ee
Therefore, the internal energy in Eq.(\ref{pvnewemr}) can be rewritten as
\begin{equation} \label{differential1}
d(M-PV)=T_h dS+ \Omega_h dJ-PdV.
\end{equation}
From Eq.(\ref{enthalpy}), we can obtain the differential relation between the enthalpy and internal energy as $dM=d U+P dV+VdP$. 
Therefore, we reach
\begin{align}
dM=TdS+ \Omega_h dJ+VdP,
\end{align}
which is  consistent with Eq.(\ref{firstlaw1}). Therefore, we conclude that as a fermion drops into the Kerr-AdS black hole, the first law of thermodynamics
is also  valid in the extended  phase space.

In the normal phase space, we have known that both the first law and second law are valid as a fermion drops into the Kerr-AdS black holes. However, this is not always true for all the cases.  The satisfaction of the first law of thermodynamics does not mean that the second law is also satisfied, especially in the extended phase space \cite{Gwak:2017kkt}. We will check the second law of  Kerr-AdS black holes in the extended phase space in the following. 
By using the Eqs.(\ref{dv}) and (\ref{pvdr}), the variation of entropy thus can be rewritten as
\bea \label{pvds}
dS=\frac{E}{\left(a^2-l^2\right)^2 \left(a^6 \left(r_h^2+l^2\right)+2 a^4 r_h^4-a^2 \left(2 l^4 r_h^2+9 l^2 r_h^4+3 r_h^6\right)+2 l^4 r_h^4\right)},
\ena
in which
\bea
E&=&2 \pi  l \left(-a^{10}   \left(r_h^2+l^2\right)dl-4 a^2 l p^r_h  \left(a^2-l^2\right)^3 \cos^2(\theta ) r_h^3+a^3  l r_h^4 \left(r_h^2+l^2\right){}^2da+4 l^7 p^r_h  r_h^5\right)\nonumber\\
&+&
2 \pi  l \left(a^6 \left(-r_h^4\right) \left(3   \left(r_h^2+l^2\right)dl+4 l p^r_h  r_h\right)-a^4 r_h^5 \left(  l^2 r_h dl+  r_h^3dl-12 l^3 p^r_h \right)\right)\nonumber\\
&+&2 \pi  l \left(-3 a^8   r_h^2 \left(r_h^2+l^2\right)dl+a^7  l \left(r_h^2+l^2\right){}^2da+2 a^5   l r_h^2 \left(r_h^2+l^2\right){}^2da
-12 a^2 l^5 p^r_h  r_h^5\right).
\ena
From Eq.(\ref{pvds}), we see that the variation of the entropy  depends on $a, l, da, dl$ while fixing $r_h$. Because we are interested  in the second laws of black holes, thus $a<l$ should be satisfied, otherwise the singularity will be naked and the thermodynamic system will break down.


\begin{figure}[h]
\centering
\includegraphics[scale=0.45]{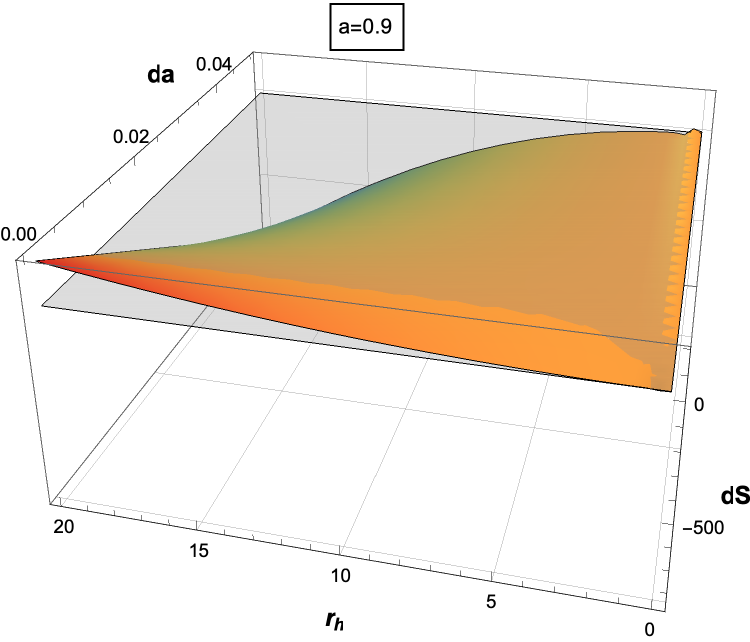}~~
\includegraphics[scale=0.45]{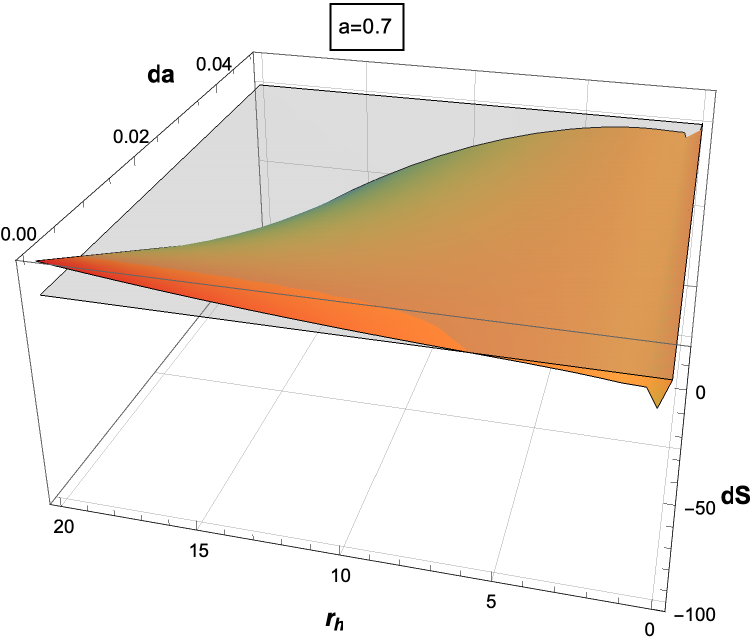}
\includegraphics[scale=0.45]{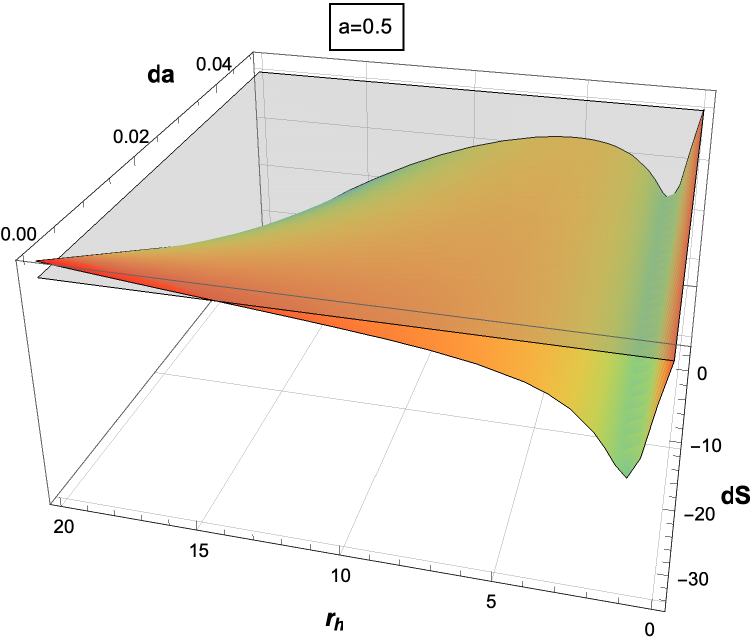}~~
\includegraphics[scale=0.45]{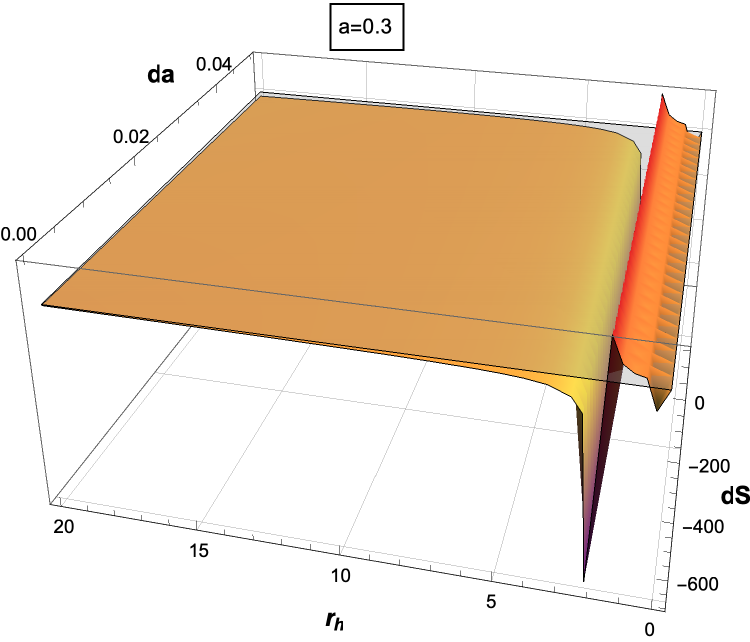}
\includegraphics[scale=0.45]{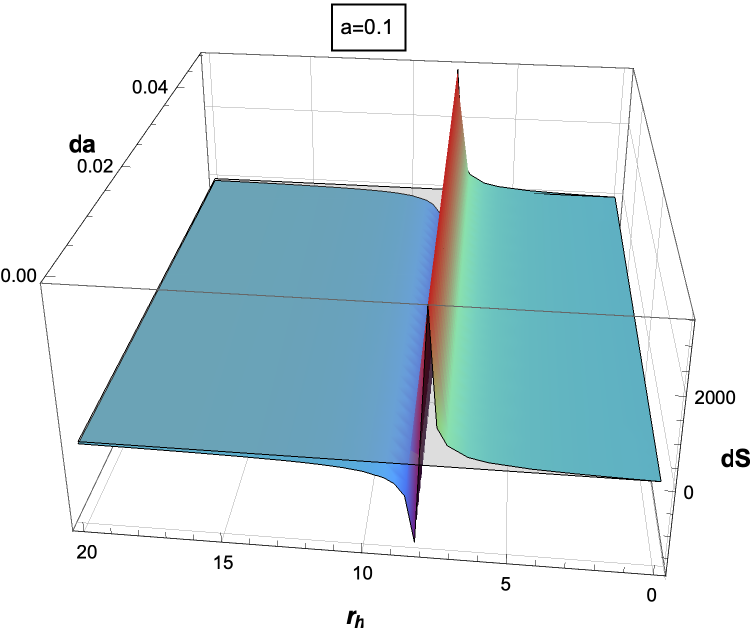}~~
\includegraphics[scale=0.45]{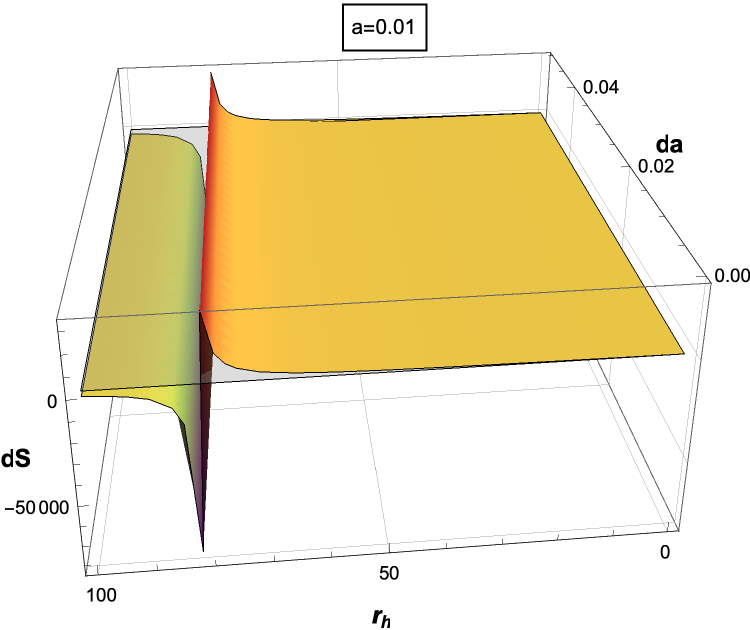}
\caption{Three dimensional visualizations of $dS$ vs. $da$ and $r_h$ for various fixed $a$'s. The horizontal gray plane is at $dS\equiv0$. We can readily see that $dS>0$ and $dS<0$ can all exist in different parameter regimes.}\label{threedplot}
\end{figure}

Since Eq.\eqref{pvds} consists of a bunch of parameters, for simplicity and without loss of generality, we will impose $p^r_h = 1, \theta = 0, dl = 0.01, l = 1$ in this subsection in order to see the effects of $da, a$ and $r_h$ to the variation of the entropy. Of course one can also set other parameters, but the general conclusions would not be changed. In Fig.\ref{threedplot} we show the 3-dimensional plots for the $dS$ in the coordinates of $da$ and $r_h$ while fixing various $a$'s. We see that $dS$ will be positive or negative for different parameter regimes. For instance of $a\equiv0.7$ and $r_h=20$ (the right panel of top row of Fig.\ref{threedplot}, there exists a critical value of $da$ such that $dS>0$ if $da<0.0067$ and $dS>0$ if $da>0.0067$. In fact, the critical values of $da$ and $r_h$ that will render $dS=0$ can be easily obtained from Eq.\eqref{pvds} as,
\begin{eqnarray}
da=\frac{a^8 \left(r_h^2+1\right)+2 a^6 r_h^2 \left(r_h^2+200 r_h+1\right)+a^4 r_h^3
   \left(r_h^3+r_h-1200\right)+1200 a^2 r_h^3-400 r_h^3}{100 a^3
   \left(r_h^2+1\right)^2 \left(a^2+r_h^2\right)}. ~~~~~~
\end{eqnarray}
As $a$ is small, for example $a\equiv0.1$ (left panel of bottom row of Fig.\ref{threedplot}), $dS>0$ if $r_h<8$ whatever $da$ is. This means the second law of thermodynamics will always be  satisfied if the black hole is near-extremal (with smaller $r_h$) independent of $da$. In the opposite limit, i.e., $r_h>8$, the second law of thermodynamics will be violated. Therefore, in the extended phase space, the second law of thermodynamics may be valid or violated, depending on the values of parameters $a, da$ and $r_h$ (given that we have fixed the values of $p^r_h, \theta, dl$ and $l$), which is different from that in the normal phase space.

\subsection{Weak  cosmic  censorship  conjecture in the extended phase space }
 Like the case in the normal phase space, we are going to  discuss the minimal value of the
function  $\Delta(r)$, since a non-positive minimum value implies the existence of  horizons while  a positive one does not. Different from that in the normal phase space,  $l$ now is a variable and the initial state is represented by $(m, a, l)$.
   As a fermion   drops into the black hole, the final state will change into $(m+dm, a+da, l+dl)$.  Correspondingly, there are also shifts for
the locations of the minimum value and event horizon, $r_l\rightarrow r_{l}+dr_{l}$, $r_h\rightarrow r_{h}+dr_{h}$. Thus, the shift for $\Delta(r)$ is
\bea \label{pvmove1}
d\Delta_{l}
= \frac{\partial \Delta_{l}}{\partial m}dm+\frac{\partial \Delta_{l}}{\partial a}da+\frac{\partial \Delta_{l}}{\partial l}dl
=-\frac{2  r_m^2 \left(a^2+r_m^2\right)dl}{l^3}+2 a  \left(\frac{r_m^2}{l^2}+1\right)da -2  r_m dm,
\ena
where we have used  $\Delta'_{l}=0$ in Eq.(\ref{condition1}). 
For the extremal black holes, we have  $r_h=r_l$ so that $\Delta_{l}=0$. In this case, Eq.(\ref{pvep}) can be used. Substituting $m$ in Eq.(\ref{ml}) and  $p^r_h$ in Eq.(\ref{pvpr}) into Eq.(\ref{pvmove1}), we obtain
\bea \label{pvdm}
dm&=&\frac{a^2 \left(-2  r_m^3  dl  - l^3{dr_m}+ l r_m^2\right){dr_m}}{2 l^3 r_m^2}
+\frac{a  \left(l^2+r_m^2\right)  da  }{l^2 r_m}+\frac{  l r_m^2 \left(l^2+3 r_m^2\right){dr_m}-2  r_m^5  dl  }{2 l^3 r_m^2}.
\ena
Inserting Eq.(\ref{pvdm}) into Eq.(\ref{pvmove1}), we get
\bea
 d\Delta_{l}=a^2  \left(\frac{1}{r_m}-\frac{r_m}{l^2}\right)dr_m+  r_m \left(-\frac{3 r_m^2}{l^2}-1\right)dr_m. \label{pvmove2}
\ena
For the extremal black holes,  Eq.(\ref{er})  is applicable. Thus, Eq.(\ref{pvmove2}) can be simplified further to be $d\Delta_{l}=0$.
Thereofre, we see that there is no shift in $\Delta_{l}$ for the extremal black holes.  The configuration of extremal black holes does not change, thus the extremal black holes are still extremal black holes as fermions are absorbed.

For the near-extremal black holes,  Eq.(\ref{er}) is not applicable at $r_{m}$ since it is valid only at $r_h$. However, we can expand it near  $r_{m}$. Replacing $r_h$  in Eq.(\ref{pvdm}) with $r_{m}+\epsilon$ and expanding it to the first order of $\epsilon$, we reach
\bea \label{pvdm2}
dM=A
+B\epsilon +O(\epsilon)^2,
\ena
in which
\bea
&&A=  \frac{a^2  {dr_m}}{2 l^2}-\frac{a^2  {dr_m}}{2 r_m^2}+\frac{a   da  }{r_m}+\frac{ {dr_m}}{2}-\frac{a^2 r_m   dl   }{l^3}+\frac{a  r_m  da  }{l^2}-\frac{ r_m^3 dl  }{l^3}+\frac{3 r_m^2{dr_m}}{2 l^2}\nonumber,\\
&&B=-\frac{a^2    dl  }{l^3}+\frac{a^2   {dr_m}}{r_m^3}+\frac{a    da  }{l^2}-\frac{a da  }{r_m^2}-\frac{3  r_m^2  dl  }{l^3}+\frac{3  r_m}{l^2}{dr_m},\nonumber
\ena
Substituting Eq.(\ref{pvdm2}) into Eq.(\ref{pvmove1}), we  obtain
\bea\label{pvdd}
d\Delta_l&=& a^2 \left(\frac{1}{r_m}-\frac{r_m}{l^2}\right) dr_m + r_m \left(-\frac{3 r_m^2}{l^2}-1\right)dr_m  \nonumber \\
&+&\frac{2  \left(a^2 \left( r_m^3 dl- l^3\right)dr_m+a  l r_m \left(l^2-r_m^2\right)da+3 r_m^4 \left( r_m dl- ldr\right) \right)}{l^3 r_m^2}\epsilon
+O(\epsilon)^2.
\ena
 Since at the horizon $\Delta(r_h)=0$, one can replace $r_h$  with $r_{m}+\epsilon$ and expand $\Delta(r_h)$ to the order of $\epsilon$,
 \be \label{pvrh}
\Delta(r_h)=a^2 \left(1-\frac{r_m^2}{l^2}\right)-\frac{3 r_m^4}{l^2}-r_m^2=0,
\ee
where we have employed  Eq.(\ref{ml}). Solving Eq.(\ref{pvrh}), we obtain
\be \label{pvl}
l= \frac{r_m \sqrt{a^2+3 r_m^2}}{\sqrt{a^2-r_m^2}}.
\ee
 Differentiating  both sides of  Eq.(\ref{pvl}), we can further get
 \be \label{pvdl}
dl= \frac{a^4  {dr_m}+6 a^2  r_m^2{dr_m} -4 a r_m^3  da  -3 r_m^4 {dr_m} }{\left(a^2-r_m^2\right){}^{3/2} \sqrt{a^2+3 r_m^2}}.
\ee
Substituting Eqs.(\ref{pvl})  and  (\ref{pvdl}) into Eq.(\ref{pvdd}), we  finally arrive at $d\Delta_l=\
O(\epsilon)^2$.
Since  $O(\epsilon)^2$ is the higher order of the small quantity $\epsilon$, thus it can be omitted. In this case, we see that the final states of the near-extremal black holes  are still the near-extremal black holes, which is similar to  that  of the extremal black holes. Therefore, we can conclude that the  weak
cosmic censorship conjecture still holds for both the extremal and near-extremal black holes in the extended phase space.

\section{Conclusions}
\label{sec5}

We investigated the laws of thermodynamics and  weak
cosmic censorship conjecture by considering a spinning fermion absorbed by the Kerr-AdS black hole. The dynamics of the fermion was investigated by the Dirac equation in the dragging coordinate system. As a result, we obtained the energy-momentum relation of the  spinning fermion near the event horizon, which was the basic of our investigation.

In the normal phase space, we  derived firstly the first law of thermodynamics by considering the  energy-momentum relation as well as the energy and  angular momentum conservation. Then we investigated the second law of thermodynamics by discussing the variation of the entropy. As predicted, it was found that the variation of the entropy is always positive. The validity of the  weak
cosmic censorship conjecture was also checked in the normal phase space by studying the variation of the function $\Delta(r)$ at the minimal point. For the case that the initial state was a black hole, the final state was  found  to be a black hole as well, thus the weak
cosmic censorship conjecture was valid as well in the Kerr-AdS black hole.

Later, we investigated the laws of thermodynamics and  weak
cosmic censorship conjecture in the extended phase space, which has not been reported previously as far as we know. In the extended phase space, the cosmological parameter is not a constant any more, however, the pressure of the thermodynamic system is considered as a constant. The first law thus contains the contribution from pressure and volume. From the  energy-momentum relation as well as the energy and  angular momentum conservation, we also derived the first law successfully. The second law, however, is more subtle since we found that it depends on the values of the spin  parameter, AdS radius as well as the their variations. As the value of the AdS radius is fixed, we found  that  the variation of entropy are different for different spin parameters. And for a fixed spin parameter, the variation of the spin parameter also affects the variation of entropy, which can be positive or negative. Therefore, the second law in the extended phase space may be valid or violated. \footnote{In the paper \cite{Hu:2019lcy}, the authors considered that the adding particles into the black hole would change the enthalpy, rather than the internal energy as we considered in our paper. This starting difference lead to their conclusion that the second law in the extended phase space would not be violated. }
We also checked the validity of the
 weak
cosmic censorship conjecture  in the extended phase space by studying the variation of the function $\Delta(r)$ at the minimal point. For the initial states were extremal and near-extremal black holes,  the final states were found to be extremal and near-extremal black holes as well. The absorbed fermion thus would not change the location of the minimum, thus the weak
cosmic censorship conjecture   is  still valid in the extended phase space.

\section*{Acknowledgements}{This work is supported  by the National
Natural Science Foundation of China (Grants No. 11675140, No. 11705005, and No. 11875095), and Basic Research Project of Science and Technology Committee of Chongqing (Grant No. cstc2018jcyjA2480).}


\end{document}